\documentclass[aps,prl,twocolumn,groupedaddress,superscriptaddress,showpacs]{revtex4-1}

\usepackage{graphicx}
\usepackage{color}
\usepackage{amsmath}
\usepackage{amsfonts}
\usepackage{amssymb}
\usepackage{dcolumn}

\begin{document}
  \title{Tricritical point in explosive percolation}

  \author{Nuno A. M. Ara\'ujo}
    \email{nuno@ethz.ch}
    \affiliation{Computational Physics for Engineering Materials, IfB, ETH Zurich, Schafmattstrasse 6, 8093 Zurich, Switzerland}

  \author{Jos\'e S. Andrade Jr}
    \email{soares@fisica.ufc.br}
    \affiliation{Computational Physics for Engineering Materials, IfB, ETH Zurich, Schafmattstrasse 6, 8093 Zurich, Switzerland}
    \affiliation{Departamento de F\'isica, Universidade Federal do Cear\'a, Campus do Pici, 60451-970 Fortaleza, Cear\'a, Brazil}

  \author{Robert M. Ziff}
    \email{rziff@umich.edu}
    \affiliation{Center for the Study of Complex Systems and Department of Chemical Engineering, University of Michigan, Ann Arbor, Michigan 48109-2136, USA}

  \author{Hans J. Herrmann}
    \email{hans@ifb.baug.ethz.ch}
    \affiliation{Computational Physics for Engineering Materials, IfB, ETH Zurich, Schafmattstr. 6, 8093 Zurich, Switzerland}
    \affiliation{Departamento de F\'isica, Universidade Federal do Cear\'a, Campus do Pici, 60451-970 Fortaleza, Cear\'a, Brazil}


  \begin{abstract}
    The suitable interpolation between classical percolation and a special variant of explosive percolation enables the explicit realization of a tricritical percolation point.
    With high-precision simulations of the order parameter and the second moment of the cluster size distribution a fully consistent tricritical scaling scenario emerges yielding the tricritical crossover exponent $1/\varphi_t=1.8\pm0.1$.
  \end{abstract}

  \maketitle

  An explicit realization of tricritical percolation has been a longstanding challenge.
  The diluted $Q$-state Potts model yields, at the tricritical point, Ising exponents, for $Q\!\!\rightarrow\!\!1$, but a direct construction of this tricritical percolation has not yet been achieved \cite{Nienhuis79,*Wu80,*Deng04}.
  The recent appearance of explosive percolation, that exhibits a discontinuous transition, provides the essential new element which enables us to find the way of establishing tricritical scaling as introduced in this Letter \cite{Achlioptas09,Ziff09,*Ziff10,Araujo10}.
  
  A discontinuous percolation transition is observed when the growth of the largest cluster is systematically suppressed \cite{Araujo10}, promoting the formation of several large components that eventually merge in an explosive way \cite{Friedman09}.
  Several aggregation models, based on percolation, have been developed to achieve this change in the nature of the transition \cite{Achlioptas09,Moreira10,Cho10,Dsouza10,Araujo10,Manna10}.
  These models are generally classified as {\it explosive percolation}, the name given in the original work that triggered the field \cite{Achlioptas09}.
  In that work, a {\it best-of-two} product rule is proposed to occupy the bond which minimizes the product of the mass of the merging clusters \cite{Ziff09,Cho09,Radicchi09,Cho10b,Radicchi10,Ziff10}.
  More recently, this procedure has been generalized to a {\it best-of-$m$} product rule in random graphs \cite{daCosta10} and regular lattices \cite{Andrade10b} to study its percolation and transport properties. 
  For different values of $m$, previously reported models are recovered \cite{Achlioptas09,Manna10,Pan10}.
  Potential applications are, for instance, the growth dynamics of the human protein homology network \cite{Rozenfeld10} and the identification of communities in real systems \cite{Pan10}.
 
  The larger the set of bonds $m$ considered in the product rule, the lower the probability that the occupied bond is related to the largest cluster \cite{Andrade10b}, promoting the compactness of the percolation cluster.
  Inspired by the $Q$-state Potts model which, in the limit $Q\!\!\rightarrow\!\!1$ its magnetic transition is related with the geometrical transition of percolation \cite{Fortuin72}, we also propose a {\it hybrid} model, where an additional parameter is included which allows us to interpolate between this discontinuous explosive regime and the one from classical percolation, characterized by a continuous transition.
  We obtain a nonequilibrium tricritical percolation where explosive percolation is diluted with classical percolation.
  Without dilution the process is explosive (discontinuous transition) while under maximum dilution classical percolation is recovered (continuous transition).
  A multicritical behavior appears due to the addition of this new degree of freedom (degree of dilution) and a new set of critical exponents is found at the tricritical point \cite{Riedel69,*Riedel72,*Riedel72b,Janssen04,*Lubeck06,*Grassberger06}.
  We also analyze the size dependence of the order parameter and the susceptibility when a {\it best-of-ten} product rule is considered, and numerically corroborate the hypothesis of a discontinuous transition.
  \begin{figure}[b]
    \includegraphics[width=\columnwidth]{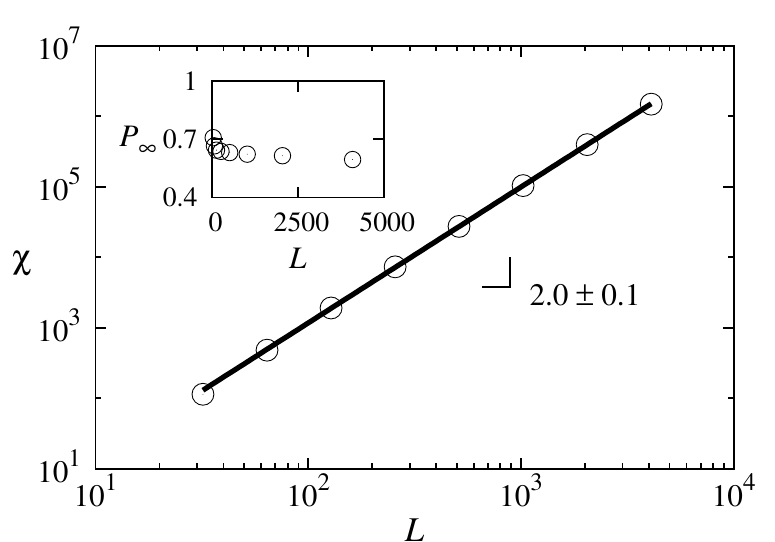}
    \caption{
      Size dependence, for the best-of-ten rule ($q=0$), of the maximum in the susceptibility $\chi$ and the order parameter $P_\infty$ (inset), at the percolation transition, for square lattices with size $L^2$, with $L$ ranging from $32$ to $4096$ sites.
                     For large system sizes, the order parameter converges to a constant value and the susceptibility scales with $L^d$, where $d$ is the dimension of the system.
                     Results have been averaged over $10^4$ samples.
                     Error bars are smaller than $0.6\%$.
      \label{fig::m10}
    }
  \end{figure}
  \begin{figure}[b]
    \includegraphics[width=\columnwidth]{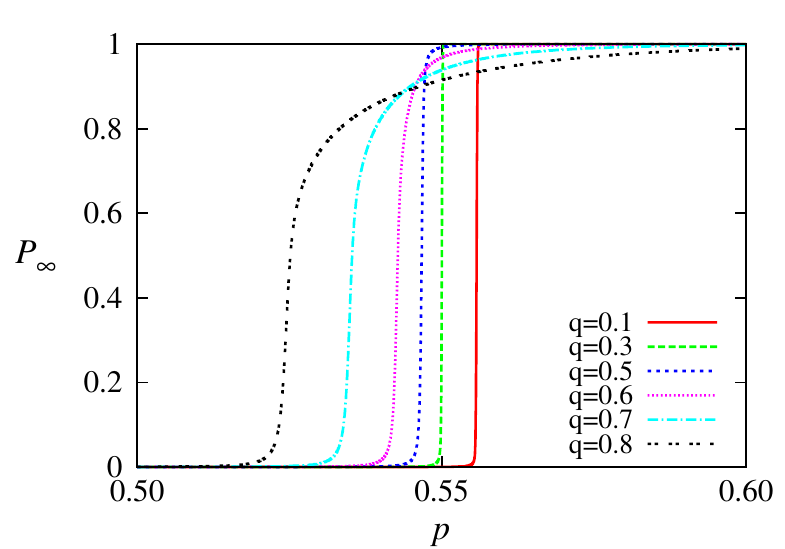}
    \caption{
      (color online) Order parameter ($P_\infty$) as a function of the fraction of occupied bonds ($p$), for different values of the degree of dilution $q$.
      Two different regimes are found.
      For lower values of $q$ the transition is discontinuous while for larger values it is continuous.
      Increasing $q$ the percolation threshold goes from the one for the best-of-ten product rule to the one for classical percolation.
      Results have been averaged over $10^4$ samples for square lattices with lateral size of $4096$, and the error bars are below $0.2\%$.
      \label{fig::largestcluster}
    }
  \end{figure}
  In the best-of-$m$ product rule, at each iteration, $m$ bonds are uniformly and randomly selected from the set of unoccupied ones.
  From this set, only the one minimizing the product of the mass of the clusters it connects is occupied.
  In the limit of $m=1$ the model recovers classical percolation \cite{Stauffer94}, while $m=2$ corresponds to the {\it Achlioptas process} \cite{Achlioptas09}.
  Nagler {\it et al.} \cite{Nagler11} have reported that for Erd\H{o}s-R\'enyi graphs (mean-field limit) a strong discontinuous transition solely occurs when $m$ is equal to the total number of bonds in the system, while for lower values of $m$ the transition is rather weakly discontinuous -- the gap shrinks with the system size and it is zero in the thermodynamic limit. 
  Also Smoluchowski equations for another model -- similar to the product rule -- reveal a continuous transition in the mean-field limit \cite{daCosta10}.
  On a square lattice, Monte-Carlo simulations reveal a discontinuous transition, characterized by some properties of a continuous one \cite{Ziff09,Radicchi10,Ziff10,Tian10}.
  Increasing the number of considered bonds $m$ promotes the formation of compact clusters, delays the percolation threshold, and, above an intermediate value, improves the conductivity of the system \cite{Andrade10b}.
  Here we take the best-of-ten case characterized by a percolation transition for the fraction of occupied bonds $p_c=0.55975\pm0.00002$, obtained from the crossing of the wrap probability for different system sizes \cite{Ziff02}.
  In Fig.~\ref{fig::m10} we show the dependence on the system size, for this model, of the maximum in the susceptibility ($\chi$) and the order parameter ($P_\infty$) at the percolation threshold.
  As in Ref.~\cite{Araujo10}, we define the susceptibility as the sum over the square of the finite cluster (i.e., nonpercolation) mass -- also known as mean cluster size -- and the order parameter as the fraction of sites belonging to the largest cluster.
  The susceptibility scales linearly with the number of sites in the lattice and the order parameter, for larger system sizes, converges to a nonzero value, as in Ref.~\cite{Radicchi10} for the best-of-two rule, consistent with the hypothesis of a discontinuous transition.
  \begin{figure}[b]
    \includegraphics[width=\columnwidth]{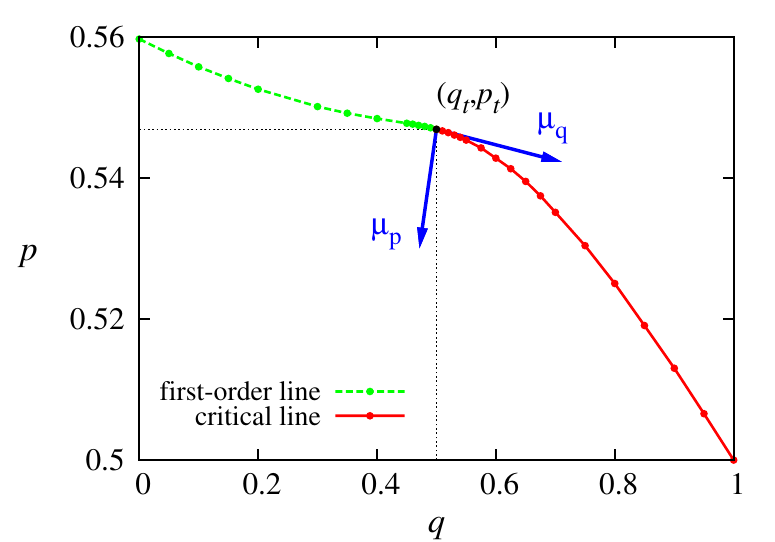}
    \caption{
      (color online) Phase diagram of the model.
                     The horizontal axis is the degree of dilution $q$ and the vertical one the fraction of occupied bonds $p$.
                     The (green) dashed line is the first-order transition line and the (red) solid line is the continuous transition.
                     Both lines meet at a tricritical point $(q_t,p_t)$.
                     Below the transition lines, there is no infinite cluster in the system.
                     Above the transition lines an infinite cluster exists.
                     The (blue) arrows stand for the relevant scaling fields $\mu_q$ and $\mu_p$ (discussed in the text).
                     The values have been obtained from the crossing of the wrap probability, with error bars below $0.2\%$.
      \label{fig::diagram}
    }
  \end{figure}
  Let us now define a hybrid model with a new parameter $q$, representing the degree of dilution.
  With probability $q$ the next occupied bond is randomly chosen among the unoccupied ones ($m=1$), while with probability $1-q$ the bond is selected following the best-of-ten product rule ($m=10$).
  Adding an additional parameter to the model, allows us to interpolate between different regimes.
  In the limit of $q=1$ we recover classical bond percolation.
  On the other limit, for $q=0$, the model boils down to the best-of-ten product rule.
  In Fig.~\ref{fig::largestcluster} we see the order parameter for different values of $q$.
  One can clearly identify two different regimes.
  While for large values of $q$ the transition is continuous, for smaller values, the transition is discontinuous.
  The larger the value of $q$ the lower the fraction of occupied bonds at the percolation threshold.
  
  The hybrid model is characterized by two control parameters, the fraction of occupied bonds $p$ and the degree of dilution $q$.
  The former is the control parameter that triggers the transition between a nonpercolative and a percolative system.
  The latter interpolates between the two different regimes, namely, the best-of-ten and the classical.
  Figure~\ref{fig::diagram} shows the two-parameter diagram of the model.
  The horizontal axis is the probability $q$ and in the vertical axis the fraction of occupied bonds $p$.
  The transition lines correspond to the values of the fraction of occupied bonds at which the percolation transition occurs for different values of $q$.
  The discontinuous line is dashed while the critical line related to a continuous transition is solid.
  All points on the critical line belong to the classical percolation universality class.
  Both lines meet at the tricritical point $(q_t,p_t)$.
  \begin{figure}[b]
    \includegraphics[width=\columnwidth]{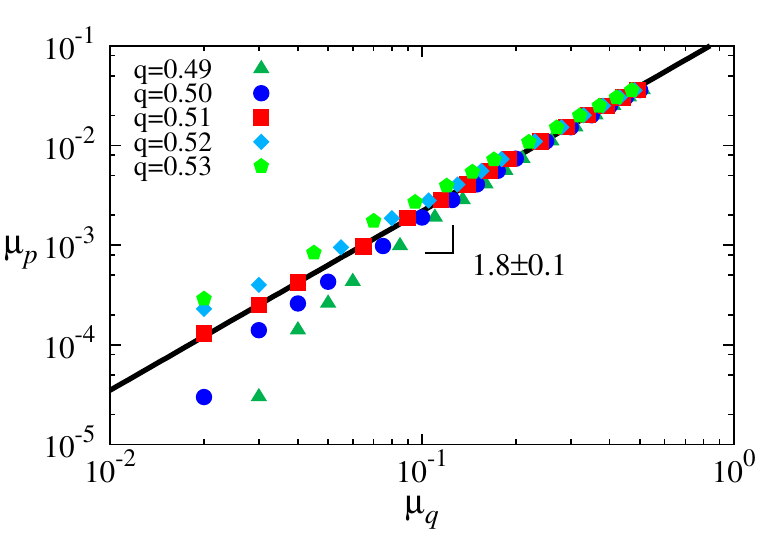}
    \caption{
      (color online) Scaling of the critical line, according to Eq.~(\ref{eq::crossoverexp}), for different values of $q$.
                     A straight line is obtained for the tricritical value $q_t=0.51\pm0.01$.
                     The obtained crossover exponent $\varphi_t$ is such that $1/\varphi_t=1.8\pm0.1$.
      \label{fig::pqscaling}
    }
  \end{figure}
  To analyze the crossover between the two different regimes we follow the theory of Riedel and Wegner \cite{Riedel69,Riedel72}.
  Two relevant scaling fields need to be defined.
  One tangent to the critical line at the tricritical point ($\mu_q$) and another perpendicular to it ($\mu_p$), as depicted in Fig.~\ref{fig::diagram}.
  In this coordinate system the critical line is described by \cite{Riedel69,Odor08}
  \begin{equation}
    \mu_p\sim\mu_q^{1/\varphi_t} \ \ , \label{eq::crossoverexp}
  \end{equation}
  \noindent where $\varphi_t$ is the tricritical crossover exponent.
  To determine the tricritical point we plot, in Fig.~\ref{fig::pqscaling}, the relation between both relevant scaling fields, along the critical line, for different values of $q$ ($0.49, 0.50, 0.51, 0.52 \mbox{ and }0.53$).
  The best straight critical line, in a log-log plot, as described by Eq.~(\ref{eq::crossoverexp}), is obtained for $q_t=0.51\pm0.01$, where the index $t$ stands for tricritical.
  For this value, the inverse of the tricritical crossover exponent is $1/\varphi_t=1.8\pm0.1$.
  The scaling fields can be obtained from the transformation
  \begin{equation}
    \left( 
      {\begin{array}{c}
        \mu_q \\
        \mu_p \\
      \end{array}}
    \right) =
    \left[ 
      {\begin{array}{cc}
        \cos\theta & -\sin\theta \\
        -\sin\theta & -\cos\theta \\
      \end{array}}
    \right]
    \left( 
      {\begin{array}{c}
        q-q_t \\
        p-p_t \\
      \end{array}}
    \right) \label{eq::rotation}
  \end{equation}
  \noindent with $\theta\approx0.022$, the angle of $\mu_p$ with the horizontal axis.
  \begin{figure}[t]
    \includegraphics[width=\columnwidth]{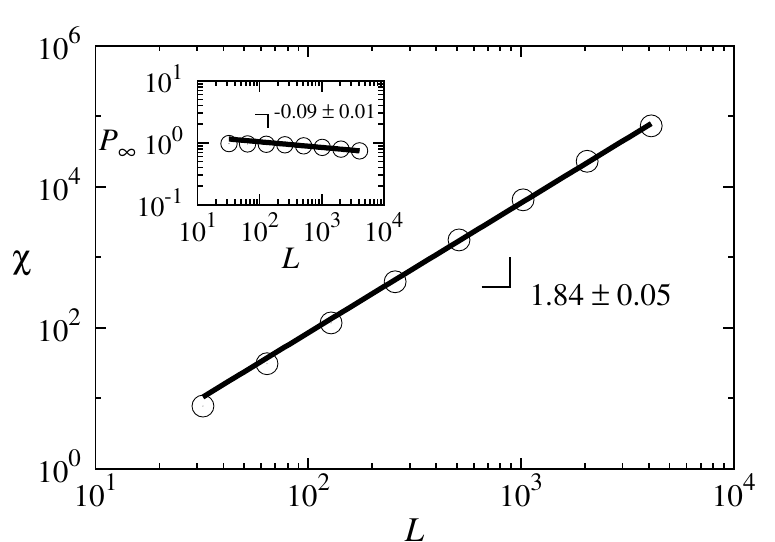}
    \caption{
      Size dependence of the maximum in the susceptibility ($\chi$) and for the order parameter ($P_\infty$), at the tricritical point of the hybrid model.
      System sizes range from $32$ to $4096$ lateral sites. 
      Results have been averaged over $10^5$ samples.
      Error bars are smaller than the point size (below $0.5\%$).
      \label{fig::tricritical}
    }
  \end{figure}
  \begin{figure}[b]
    \includegraphics[width=\columnwidth]{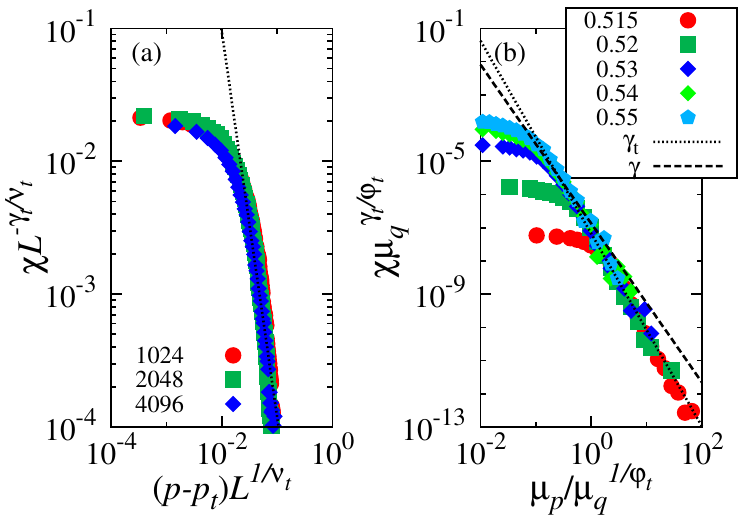}
    \caption{
      (color online) (a) Finite-size scaling for the susceptibility ($\chi$) for $q=q_t$, according to the scaling {\it ansatz} of Eq.~\ref{eq::scalings}.
                     Results for square lattices with lateral size of $1024$, $2048$, and $4096$ sites, are averaged over $10^4$ samples.
                     The value of $\gamma_t=2.9\pm0.1$ can be obtained from the slope of the dotted line.
                     (b) Crossover scaling for the susceptibility ($\chi$) for different values of $q=\{0.515,0.52,0.53,0.54,0.55\}$ and the largest system size ($4096^2$ sites).
                         Dashed and dotted lines correspond to the two different regimes (details in the text).
                         Results have been averaged over $10^4$ samples of a square lattice with $4096$ sites of lateral size.
      \label{fig::scalings}
    }
  \end{figure}
  For the typical percolation transition obtained by crossing the critical line in the direction of the $\mu_p$ scaling field, the tricritical point is characterized by a new set of tricritical exponents -- not only the crossover exponent $\varphi_t$, but also, for example, $\beta_t$, $\gamma_t$, and $\nu_t$.
  To obtain these exponents we consider, in Fig.~\ref{fig::tricritical}, the size dependence of both the maximum in the susceptibility ($\chi$) and the order parameter ($P_\infty$), at the tricritical point.
  The former gives $\gamma_t/\nu_t=1.84\pm0.05$ and the latter $\beta_t/\nu_t=0.09\pm0.01$.
  The scaling {\it ansatz} for the susceptibility is given by
  \begin{equation}\label{eq::scalings}
    \chi\left(p-p_t,L\right)=L^{\gamma_t/\nu_t}F\left[(p-p_t)L^{1/\nu_t}\right] \ \ ,
  \end{equation}
  \noindent where $F\left[x\right]$ is a scaling function which, for large $x$ scales as $x^{-\gamma_t}$.
  Figure~\ref{fig::scalings}(a) shows the scaling for three different system sizes, namely, $1024$, $2048$, and $4096$, giving $\gamma_t=2.9\pm0.1$.
  Also, from Figs.~\ref{fig::tricritical}~and~\ref{fig::scalings}(a), we find $\nu_t=1.56\pm0.09$ and, consequently, $\beta_t=0.14\pm0.03$.

  For the complete tricritical crossover scaling of the susceptibility, $\chi$ is considered as a homogeneous function on the relevant scaling fields,
  \begin{equation}
    \chi\left(l\mu_p,l^{\varphi_t}\mu_q,l^{1/\delta_t} h\right)=l^{-\gamma_t}\chi\left(\mu_p,\mu_q,h\right) \ \ , \label{eq::homogeneous}
  \end{equation}
  \noindent where $l$ is a scaling parameter, $h$ is the ghost field of percolation, and $\delta_t$ can be obtained from $\delta_t=(\gamma_t+\beta_t)/\beta_t$ \cite{Gaunt76,*Fisher74,*Stauffer79}.
  In the absence of ghost field ($h=0$) and with $l=\mu_q^{-1/\varphi_t}$ one has,
  \begin{equation}
    \chi\left(\mu_p,\mu_q\right)=\mu_q^{-\gamma_t/\varphi_t}F_\mathrm{cross}\left[\mu_p/\mu_q^{1/\varphi_t}\right] \ \ , \label{eq::scaling}
  \end{equation}
  \noindent where $F_\mathrm{cross}\left[y\right]$ is a scaling function such that, $F_\mathrm{cross}\left[y\right]\sim y^{-\gamma}$ for $y\ll1$ and $F_\mathrm{cross}\left[y\right]\sim y^{-\gamma_t}$ for $y\gg1$ \cite{Lubeck06}.
  Figure~\ref{fig::scalings}(b) shows the tricritical crossover scaling for different values of $q$, close to the tricritical point.
  Two different regimes are clearly identified.
  For low values of $\mu_p/\mu_q^{1/\varphi_t}$ (below $1$), corresponding to the larger values of $q$, results are consistent with $\gamma=43/18$ as expected for the classical percolation universality class (dashed line).
  For large values, the exponent of the scaling function corresponds to the one from the tricritical point reported in this work.
  This scaling gives the tricritical crossover between the discontinuous (explosive) percolation and the continuous one.

  Our model can be further generalized to different values of $m$ (number of preselected bonds).
  As discussed previously, the larger the value of $m$ the stronger the signs of a discontinuous transition for the pure best-of-$m$ model \cite{Andrade10b}.
  As a consequence, the level of dilution at the tricritical point $q_t$ must increase with $m$.

  In short, we have introduced a diluted tricritical percolation model, where the degree of dilution allows us to interpolate between the classical percolation regime, characterized by a continuous transition, and the best-of-ten one where, on a square lattice, a discontinuous transition is observed.
  We have presented a two-parameter diagram for the model to systematize the different percolation regimes.
  Two transition lines were identified: a discontinuous and a critical line.
  Both lines meet at a tricritical point characterized by a new set of tricritical exponents different from the ones of the classical percolation universality class.
  Finally, we also analyze the tricritical crossover scaling for the susceptibility.

\begin{acknowledgments}
We acknowledge a helpful discussion with D. P. Landau, and financial support from the ETH Competence Center Coping with Crises in Complex Socio-Economic Systems (CCSS) through ETH Research Grant No. CH1-01-08-2.
We also acknowledge the Brazilian agencies CNPq, CAPES and FUNCAP, and the grant CNPq/FUNCAP, for financial support.
\end{acknowledgments}

\bibliography{tricritical}

\end{document}